\documentclass[12pt]{article}
\makeatletter
\def\eqnarray{%
 \stepcounter{equation}%
 \let\@currentlabel=\theequation
 \global\@eqnswtrue
 \global\@eqcnt\z@
 \tabskip\@centering
 \let\\=\@eqncr
 $$\halign to \displaywidth\bgroup\@eqnsel\hskip\@centering
 $\displaystyle\tabskip\z@{##}$&\global\@eqcnt\@ne
 \hfil$\displaystyle{{}##{}}$\hfil
 &\global\@eqcnt\tw@$\displaystyle\tabskip\z@{##}$\hfil
 \tabskip\@centering&\llap{##}\tabskip\z@\cr}
\makeatother

\newcommand{\psu}{PSU(2$|$2)}
 
\newcommand{\U}{$\otimes$U(1)$^3$}
\newcommand{\D}{D(2,1;$\gamma$)}
\newcommand{\bb}{\begin{eqnarray}}
\newcommand{\ee}{\end{eqnarray}}
\newcommand{\oo}{$\otimes$}

\begin{document}

\title{ Spin-chain with PSU(2$|$2)$\otimes$U(1)$^3$ \\
  and  Non-linear $\sigma$-model with \D }

\author{Shogo Aoyama\thanks{e-mail: spsaoya@ipc.shizuoka.ac.jp}
  \hspace{1cm}  Yuco Honda \thanks{e-mail: yufrau@hotmail.co.jp } \\
       Department of Physics \\
              Shizuoka University \\
                Ohya 836, Shizuoka  \\
                 Japan}

\date{}
\maketitle
\begin{abstract}
We propose that the spin-chain with the PSU(2$|$2)$\otimes$U(1)$^3$ symmetry 
is equivalent to the non-linear $\sigma$-model on PSU(2$|$2)$\otimes$U(1)$^3$/\{H$\otimes$U(1)\} 
 with a certain subgroup H. To this end we show that the spin-variable of the former theory is identified as the Killing scalar of the latter and their correlation functions can have   the same integrability. It is crucial to think  that the respective theory gets the PSU(2$|$2)$\otimes$U(1)$^3$ symmetry by a symmetry reduction  the exceptional supergroup \D, rather than  by an extension of PSU(2$|$2).  

\end{abstract}

\vspace{5cm}

\noindent
Keywords

Non-linear $\sigma$-model, Spin-chain, Supergroup, R-matrix

\newpage

\section{Introduction}

The study of the string/QCD duality has a long history going back to the late 1970s. During the last decade the subject has been studied with a renewed interest. The most clear-cut assertion of the string/QCD duality was made by calculating the anomalous dimensions of a spin-chain system on one side and the ${\cal N}=4$ SUSY QCD on the other and showing a remarkable agreement between them\cite{Ma}.  A spin-chain system of variables $\psi_m(x)$ was defined by correlation functions taking the form\cite{Bei} 
\bb
e^{ip_1x_1+ip_2x_2+\cdots\cdots}<\psi_{m_1}(x_1)\psi_{m_2}(x_2)\cdots\cdots>. \label{cor}
\ee
Here $\psi_m(x)$ is assumed to be the fundamental vector of the \psu\U\  symmetry.
 It is also assumed that the correlation functions obey the exchange algebra with the R-matrix for two adjacent variables $\psi_m(x)$s. If the symmetry of the spin-system were strictly \psu, the R-matrix would be universally given by the plug-in formula for \psu\cite{Pres,Kho}, and the correlation functions would be position-independent as those of the topological theory\cite{Ao1}. The \psu\ symmetry had to be centrally extended to \psu\U\ in order to give an account of the duality  to the ${\cal N}=4$ SUSY QCD. A position-dependent R-matrix was found in an explicit form\cite{Bei}, which is not of the difference form of the two spectral parameters. This unusual feature of the R-matrix attracted a vivid interest among the community of mathematical physics\cite{Bei2}. A keen insight into the matter was given by realizing \psu\U\ as a symmetry reduction of the exceptional group \D\cite{Bei,Mo}. 

The above arguments are  based on the integrability  and the \psu\U\ symmetry. 
They are just assumed and their origin is obscure. These assumptions get a firm base by considering a 2-d non-linear $\sigma$-model with the \psu\U\ symmetry as an equivalent theory to the spin-system. Namely the non-linear $\sigma$-model is integrable admitting an infinite number of conserved currents. The \psu\U\  symmetry  is regarded as descending from the superconformal symmetry of  the IIB superstring.

According to \cite{Ao2} a quantity corresponding to the spin-variable $\psi_m$ with the \psu\ symmetry  may be constructed as the Killing scalar on a coset space \psu/H with a certain subgroup.\footnote{The Killing scalar was discussed originally as a quantity called the G-primary  in the constrained WZWN model on the coset space G/\{H$\otimes$U(1)$^d$\}\cite{Ao3,Ao4}. In the context of the non-linear $\sigma$-model it was discussed in \cite{Ao6}} The exchange algebra for the Killing scalar may be discussed by studying the Poisson structure in the non-linear $\sigma$-model on \psu/H.  The R-matrix of the exchange algebra is universal following the quantization of \cite{Ao2}. 

The aim of this letter is to show that  when the spin-system has  the centrally extended symmetry \psu\U\ \cite{Bei}, the equivalent theory is  
 the non-linear $\sigma$-model on an enlarged coset space \psu\U/\{H$\otimes$U(1)\}. That is, we show that  the Killing scalar of this generalized non-linear $\sigma$-model has the same transformation property as the spin-variable $\psi_m$  with the \psu\U\ symmetry. 
 As the result the non-linear $\sigma$-model gets the position-dependent
  R-matrix of the spin-system and  correlation functions taking the centrally extended form identical to (\ref{cor}).  
But the coset space \psu\U/\{H$\otimes$U(1)\} is meaningless as it is, because the ordinary non-linear realization is not applicable to a non-simple group such as \psu\U. To give it a precise meaning   we consider the non-linear $\sigma$-model on  a further enlarged coset space \D/\{H$\otimes$U(1)\}. 
The non-linear $\sigma$-model on \psu\U/\{H$\otimes$U(1)\} is defined by the symmetry reduction of \D\ to \psu\U\ in the model on this enlarged coset space. The symmetry reduction is undertaken in the same way as was done  for the spin-system\cite{Bei,Mo}.

This letter is organized as follows. In section 2 we explain the Lie-algebra of \D\ and  a reducing process to the subgroups \psu\U, SU(2$|$2), PSU(2$|$2) at an algebraic level. In section 3 the matrix representation for those subgroups are given. The non-linear representation of \D\ is discussed in section 4. In section 5 we undertake the symmetry reduction, discussed at the algebraic level in section 2, in the coset space \D/\{H$\otimes$U(1)$^3$\}. It is then shwon that  the non-linear $\sigma$-model on the reduced coset space is equivalent to the spin-chain with the \psu\ symmetry. 

\section{The Lie-algebra of \D}

The Lie-algebra of \D\ is given by 17 generators
\bb
\{T_{\rm D}^\Xi\}=\{F^{\alpha a\dot\alpha},L^\alpha_{\ \beta},R^a_{\ b},{\dot L}^{\dot\alpha}_{\ \dot\beta}\}.  \label{TD}
\ee
$L^\alpha_{\ \beta},R^a_{\ b},{\dot L}^{\dot\alpha}_{\ \dot\beta}$ generate the subgroup SU(2)\oo SU(2)\oo SU(2) and   $F^{\alpha a\dot\alpha}$ are fermionic generators which enlarge  the subgroup to \D. The Lie-algebra takes the form\cite{Bei,Mo} 
\bb
&\ &[L^\alpha_{\ \beta},L^\gamma_{\ \delta}]=-\delta^\gamma_\beta L^\alpha_{\ \delta}+ \delta^\alpha_\delta L^\gamma_{\ \beta}, \quad\quad \label{3}\\ 
&\ &[R^a_{\ b},R^c_{\ d}]=-\delta^c_b R^a_{\ d}+ \delta^a_d R^c_{\ b}, \label{3'}\\
&\ & [{\dot L}^{\dot\alpha}_{\ \dot\beta},{\dot L}^{\dot\gamma}_{\ \dot\delta}]=-\delta^{\dot\gamma}_{\dot\beta} {\dot L}^{\dot\alpha}_{\ \dot\delta}+ \delta^{\dot\alpha}_{\dot\delta} {\dot L}^{\dot\gamma}_{\ \dot\beta}, \label{4} \\
&\ & [L^\alpha_{\ \beta}, F^{\gamma c\dot \gamma}]=-\delta^\gamma_\beta F^{\alpha c\dot \gamma}+{1\over 2}\delta^\alpha_\beta  F^{\gamma c\dot \gamma}, \quad\quad \label{5}\\
&\ &[R^a_{\ b}, F^{\gamma c\dot \gamma}]=-\delta^c_b F^{\gamma a\dot \gamma}+{1\over 2}\delta^a_b  F^{\gamma c\dot \gamma}, \label{5'} \\
&\ & [{\dot L}^{\dot\alpha}_{\ \dot\beta}, F^{\gamma c\dot \gamma}]=-\delta^{\dot\gamma}_{\dot\beta} F^{\gamma c\dot\alpha}+{1\over 2}\delta^{\dot\alpha}_{\dot\beta}  F^{\gamma c\dot \gamma}, \label{6} \\
&\ & \{ F^{\alpha a\dot\alpha},F^{\beta b\dot\beta} \}=\alpha\epsilon^{\alpha\kappa}\epsilon^{ab}\epsilon^{\dot\alpha\dot\beta}L^\beta_{\ \kappa}+ \beta\epsilon^{\alpha\beta}\epsilon^{a k}\epsilon^{\dot\alpha\dot\beta}R^b_{\ k} +\gamma \epsilon^{\alpha\beta}\epsilon^{ab}\epsilon^{\dot\alpha\dot\kappa}{\dot L}^{\dot\beta}_{\ \dot\kappa}, \label{7}
\ee
with $\alpha+\beta+\gamma=0$. The overall scaling does not change the algebraic structure. The algebra is characterized by the only parameter ${\gamma\over \alpha}$. With the definition 
\bb
{\dot L}^{\dot\alpha}_{\ \dot\beta}=\left[
\begin{array}{c|c}
C &  K\\
\hline
-P & -C
\end{array}\right]^{\dot\alpha}_{\ \dot\beta}\ \ , \quad\quad\quad
\Bigg[F^{\alpha a}\Bigg]^{\dot\alpha}
=
 \left[
\begin{array}{c}
\epsilon^{ak}S^\alpha_{\ k}  \\
\hline
 \epsilon^{\alpha \kappa}Q^a_{\ \kappa}
\end{array}\right]^{\dot\alpha},    \nonumber
\ee
(\ref{TD}) is decomposed as 
\bb
\{T_{\rm D}^\Xi\}=\{Q^a_{\ \alpha},S^\alpha_{\ a},L^\alpha_{\ \beta},R^a_{\ b},C,P,K\}.  \label{TD'}
\ee
In this base the algebrae (\ref{4}), (\ref{6}) and (\ref{7}) become
\bb
&\ &\{Q^a_{\ \alpha},Q^b_{\ \beta}\}=\gamma\epsilon_{\alpha\beta}\epsilon^{ab}P, \quad\quad 
\{S^\alpha_{\ a},S^\beta_{\ b}\}=\gamma\epsilon^{\alpha\beta}\epsilon_{ab}K,
 \label{9}\\
&\ & \{Q^a_{\ \alpha},S^\beta_{\ b} \}=\alpha\delta^a_b L^\beta_{\ \alpha}
 -\beta\delta^\beta_\alpha R^a_{\ b} +\gamma\delta_\alpha^\beta\delta_b^a C,
   \label{10}\\
&\ & [C,P]= P, \quad\quad [C,K]= -K, \quad\quad [P,K]=-2C,   \label{11}\\
&\ & [C, S^\alpha_{\ a}]=-{1\over 2}S^\alpha_{\ a}, \quad\quad
[P,S^\alpha_{\ a}]=-\epsilon^{\alpha\beta}\epsilon_{ab}Q^b_{\ \beta},
\quad\quad \\
&\ &[K,S^\alpha_{\ a}]=0, \label{12}\\
&\ & [C, Q^a_{\ \alpha}]={1\over 2}Q^a_{\ \alpha}, \quad\quad
[P, Q^a_{\ \alpha}]=0, \quad\quad \\
&\ &[K,Q^a_{\ \alpha}]=
  \epsilon_{\alpha\beta}\epsilon^{ab}S^\beta_{\ b},  \label{13}
\ee
and the algebrae (\ref{3}), (\ref{3'}), (\ref{5}), (\ref{5'}) do not change the forms, i.e., 
\bb
&\ &[L^\alpha_{\ \beta},L^\gamma_{\ \delta}]=-\delta^\gamma_\beta L^\alpha_{\ \delta}+ \delta^\alpha_\delta L^\gamma_{\ \beta}, \quad\quad\quad 
[R^a_{\ b},R^c_{\ d}]=-\delta^c_b R^a_{\ d}+ \delta^a_d R^c_{\ b}, \nonumber\\
&\ & [L^\alpha_{\ \beta}, S^\gamma_{\ c}]=-\delta^\gamma_\beta S^\alpha_{\ c} +{1\over 2}\delta^\alpha_\beta S^\gamma_{\ c}, \quad\quad\ \  [R^a_{\ b}, Q^c_{\ \gamma}]=-\delta^c_b Q^a_{\ \gamma}+{1\over 2}\delta^a_b Q^c_{\ \gamma}, \nonumber\\
&\ & [L^\alpha_{\ \beta}, Q^c_{\ \gamma}]=\delta^\alpha_\gamma Q^c_{\ \beta} -{1\over 2}\delta^\alpha_\beta Q^c_{\ \gamma}, \quad\quad\ \  [R^a_{\ b}, S^\gamma_{\ c}]=\delta^a_c S^\gamma_{\ b}-{1\over 2}\delta^a_b S^\gamma_{\ c}, 
\label{14} 
\ee
The quadratic Casimir is given by 
\bb
&\ & T=\alpha L^\alpha_{\ \beta}L^\beta_{\ \alpha}+\beta R^a_{\ b}R^b_{\ a}
 +\gamma {\dot L}^{\dot\alpha}_{\ \dot\beta}{\dot L}^{\dot\beta}_{\ \dot\alpha}
 -\epsilon_{\alpha\beta}\epsilon_{ab}\epsilon_{\dot\alpha\dot\beta}F^{\alpha a\dot\alpha}F^{\beta b\dot\beta}    \nonumber \\
&\ & \hspace{1cm}
 =\alpha L^\alpha_{\ \beta}L^\beta_{\ \alpha}+\beta R^a_{\ b}R^b_{\ a}
 +\gamma(2C^2-PK-KP)+S^\alpha_{\ a}Q^a_{\ \alpha}-Q^a_{\ \alpha}S^\alpha_{\ a},     \nonumber
\ee
in the respective basis of the algebrae (\ref{3})$\sim$(\ref{7}) and (\ref{9})$\sim$(\ref{14}).

To obtain the Lie-algebra of \psu\U, we rescale 
as $(C,P,K)\rightarrow {1\over \gamma}(C,P,K)$\cite{Bei,Mo}. In the limit $\gamma\rightarrow 0$\footnote {In \cite{Bei,Mo} they took the limit $\epsilon\rightarrow 0$ for $\alpha=1,\beta=-1-\epsilon,\gamma=\epsilon$. It is the same as our limit by overall scaling. } (\ref{9}) and (\ref{10}) become 
\bb
&\ &\{Q^a_{\ \alpha},Q^b_{\ \beta}\}=\epsilon_{\alpha\beta}\epsilon^{ab}P, \quad\quad 
\{S^\alpha_{\ a},S^\beta_{\ b}\}=\epsilon^{\alpha\beta}\epsilon_{ab}K,
 \nonumber\\
&\ & \{Q^a_{\ \alpha},S^\beta_{\ b} \}=\alpha\delta^a_b L^\beta_{\ \alpha}
 -\beta\delta^\beta_\alpha R^a_{\ b} +\delta_\alpha^\beta\delta_b^a C,
   \label{16}
\ee
with $\alpha+\beta=0$, 
while (\ref{11})$\sim$(\ref{13}) vanishing algebrae. Together with the algebrae (\ref{14}) they are closed to the centrally extended algebra of \psu\U. We denote the generators of this algebra by 
\bb
\{T_{{\rm PSU}\otimes{\rm U(1)}^3 }^\Xi\}=\{Q^a_{\ \alpha},S^\alpha_{\ a},L^\alpha_{\ \beta},R^a_{\ b},C,P,K\}.  \label{17}
\ee
Here the generators are the same one as given by (\ref{TD'}), but by the above rescaling the SU(2) symmetry given by (\ref{11}) is broken to U(1)$^3$. We shall 
call this symmetry reduction the \psu\U\ limit. Actually  the algebrae (\ref{14}) and (\ref{16}) are closed to even smaller algebrae, of which generators are 
\bb
\{T_{{\rm PSU}\otimes{\rm U(1)}^3 }^\Xi\}\Big|_{P=K=0}=\{T^\Xi_{\rm SU}\}=\{Q^a_{\ \alpha},S^\alpha_{\ a},L^\alpha_{\ \beta},R^a_{\ b},C\},  \label{18}
\ee
and 
\bb
\{T_{{\rm PSU}\otimes{\rm U(1)}^3 }^\Xi\}\Big|_{C=P=K=0}=\{T^\Xi_{\rm PSU}\}=\{Q^a_{\ \alpha},S^\alpha_{\ a},L^\alpha_{\ \beta},R^a_{\ b}\}.  \label{19}
\ee
The symmetry reductions are called the SU(2$|$2) and PSU(2$|$2) limits respectively.

\section{Matrix representation}

The generators of \psu\U\ in (\ref{14}) and (\ref{16}) can be represented by 4$\times$4 matrices\cite{Bei,Mo} as
\bb
 \scriptstyle{L^\alpha_{\ \beta}}=\left(
\begin{array}{c|c}
\vspace{-0.3cm}  &    \\
\hspace{-0.1cm}\scriptstyle{\delta^\alpha_\delta\delta^\gamma_\beta-{1\over 2}\delta^\alpha_\beta\delta^\gamma_\delta}  &   \hspace{0.7cm} \scriptstyle{0} \hspace{0.7cm}    \\
\vspace{-0.2cm}  &    \\
  \hline   
\vspace{-0.3cm}  &    \\
\scriptstyle{0 }  & \scriptstyle{0} 
\vspace{-0.3cm} \\
  &     \\
\end{array}\right),\quad\quad\ 
 \scriptstyle{R^a_{\ b}}=\left(
\begin{array}{c|c}
\vspace{-0.3cm}  &    \\
\hspace{0.7cm}\scriptstyle{0}\hspace{0.7cm}  &   \hspace{0.7cm} \scriptstyle{0} \hspace{0.7cm}    \\
\vspace{-0.2cm}  &    \\
  \hline   
\vspace{-0.3cm}  &    \\
\scriptstyle{0 }  & \scriptstyle{\delta^a_d\delta^c_b-{1\over 2}\delta^a_b\delta^c_d}  
\vspace{-0.3cm} \\
   &     \\
\end{array}\right),    \nonumber
\ee
\vspace{-0.3cm} 
\bb
\scriptstyle{Q^a_{\ \alpha}}=\left(
\begin{array}{c|c}
\vspace{-0.3cm}  &    \\
\hspace{0.73cm} \scriptstyle{0} \hspace{0.73cm}    & \scriptstyle{{\cal A}\delta^\gamma_\alpha\delta^a_d}   \\
\vspace{-0.2cm}  &    \\
  \hline   
\vspace{-0.3cm}  &    \\
\scriptstyle{{\cal B}\epsilon^{ac}\epsilon_{\alpha\delta}}  & \hspace{0.73cm} \scriptstyle{0} \hspace{0.73cm}  
\vspace{-0.3cm} \\
  &     \\
\end{array}\right),\quad\quad
\scriptstyle{S^\alpha_{\ a}}=\left(
\begin{array}{c|c}
\vspace{-0.3cm}  &    \\
\hspace{0.68cm} \scriptstyle{0} \hspace{0.68cm}    & \scriptstyle{{\cal C}\epsilon_{ad}\epsilon^{\alpha\gamma}}  \\
\vspace{-0.2cm}  &    \\
  \hline   
\vspace{-0.3cm}  &    \\
\scriptstyle{{\cal D}\delta^\alpha_\delta\delta^c_a}  & \hspace{0.75cm} \scriptstyle{0} \hspace{0.75cm}  
\vspace{-0.3cm} \\
  &     \\
\end{array}\right),  \nonumber
\ee
\vspace{-0.3cm}
\bb
\ \ \scriptstyle{P}=\left(
\begin{array}{c|c}
\vspace{-0.3cm}  &    \\
\hspace{-0.1cm}\scriptstyle{{\cal AB}\delta^\gamma_\delta}  &   \hspace{0.7cm} \scriptstyle{0} \hspace{0.7cm}    \\
\vspace{-0.2cm}  &    \\
  \hline   
\vspace{-0.3cm}  &    \\
  & \scriptstyle{{\cal AB}\delta^c_d} 
\vspace{-0.3cm} \\
 \hspace{0.73cm} \scriptstyle{0} \hspace{0.73cm}  &     \\
\end{array}\right),\quad\quad\ \ 
 \scriptstyle{K}=\left(
\begin{array}{c|c}
\vspace{-0.3cm}  &    \\
\hspace{-0.1cm}\scriptstyle{{\cal CD}\delta^\gamma_\delta}  &   \hspace{0.7cm} \scriptstyle{0} \hspace{0.7cm}    \\
\vspace{-0.2cm}  &    \\
  \hline   
\vspace{-0.3cm}  &    \\
  & \scriptstyle{{\cal CD}\delta^c_d} 
\vspace{-0.3cm} \\
 \hspace{0.73cm} \scriptstyle{0} \hspace{0.73cm}  &     \\
\end{array}\right),  \nonumber
\ee
\vspace{-0.3cm}
\bb
\scriptstyle{C}=\left(
\begin{array}{c|c}
\vspace{-0.3cm}  &    \\
\hspace{-0.15cm}\scriptstyle{{1\over 2}({\cal AD+BC})\delta^\gamma_\delta}\hspace{-0.05cm}  &   \hspace{0.7cm} \scriptstyle{0} \hspace{0.7cm}    \\
\vspace{-0.2cm}  &    \\
  \hline   
\vspace{-0.3cm}  &    \\
  & \hspace{-0.05cm}\scriptstyle{{1\over 2}({\cal AD+BC})\delta^c_d} \hspace{-0.15cm}
\vspace{-0.3cm} \\
 \hspace{0.73cm} \scriptstyle{0} \hspace{0.73cm}  &     \\
\end{array}\right),    \label{20}
\ee
with the index notation of a 4$\times$4 supermatrix
$$
t=\left(
\begin{array}{c|c}
\vspace{-0.3cm} & \\
 \hspace{0.55cm} t^\gamma_{\ \delta}\hspace{0.55cm} &  t^\gamma_{\ d}  \\
\vspace{-0.2cm} & \\
\hline
\vspace{-0.3cm} & \\
 t^c_{\ \delta}   & \hspace{0.55cm} t^c_{\ d} \hspace{0.55cm} 
\vspace{-0.3cm}  \\
 &   
\end{array}\right). \ \ 
$$
Here the constants ${\cal A,B,C,D}$ are constrained by
\bb
{\cal AD}-{\cal BC}=\alpha=-\beta.    \label{21} 
\ee

Let $\psi$ to be a linear representation vector of \psu\U. It transforms as
\bb
\delta\psi=i\epsilon\cdot T_{ {\rm PSU}\otimes{\rm U(1)}^3} \ \psi \equiv \left(
\begin{array}{c|c}
\vspace{0cm} & \\
 {\epsilon_L}^\gamma_{\ \delta}-{1\over 2}\delta^\gamma_\delta\epsilon_L+\delta^\gamma_\delta c& {\cal A}{\epsilon_Q}^\gamma_{\ d}+{\cal C}{\epsilon_S}^b_{\ \beta}\epsilon^{\beta\gamma}\epsilon_{bd} \\
\vspace{0cm} & \\
\hline
\vspace{0cm} & \\
 {\cal D}{\epsilon_S}^c_{\ \delta}+{\cal B}{\epsilon_Q}^\alpha_{\ a}\epsilon^{ac}\epsilon_{\alpha\delta}   & {\epsilon_R}^c_{\ d}-{1\over 2}\delta^c_d\epsilon_R + \delta^c_d c
\vspace{0cm}  \\
 &   
\end{array}\right)\ \psi,   \label{22}
\ee
with 
$$
c={1\over 2}\epsilon_C({\cal AD}+{\cal BC})+\epsilon_P {\cal AB}+\epsilon_K {\cal CD}
$$
Here the generators $T^\Xi_{{\rm PSU}\otimes{\rm U(1)}^3}$ were given in (\ref{17}), and use was made of the corresponding infinitesimal parameters 
$$
\{\epsilon^\Xi\}=\{{\epsilon_L}^\alpha_{\ \beta},{\epsilon_R}^a_{\ b}, {\epsilon_Q}^\alpha_{\ a},{\epsilon_S}^a_{\ \alpha},\epsilon_C,\epsilon_P, \epsilon_K\}.
$$
When ${\cal B}={\cal C}=0$ (\ref{22}) becomes the linear transformation of SU(2$|$2)(=\psu$\otimes$U(1))
\bb
\delta\psi=i\epsilon\cdot T_{\rm SU} \ \psi \equiv \left(
\begin{array}{c|c}
\vspace{0cm} & \\
 {\epsilon_L}^\gamma_{\ \delta}-{1\over 2}\delta^\gamma_\delta\epsilon_L+{1\over 2}\delta^\gamma_\delta \epsilon_C  & {\cal A}{\epsilon_Q}^\gamma_{\ d} \\
\vspace{0cm} & \\
\hline
\vspace{0cm} & \\
 {\cal D}{\epsilon_S}^c_{\ \delta}   & {\epsilon_R}^c_{\ d}-{1\over 2}\delta^c_d\epsilon_R + {1\over 2}\delta^c_d \epsilon_C
\vspace{0cm}  \\
 &   
\end{array}\right)\ \psi,    \label{23}
\ee
with the constraints (\ref{21}) simplified to be ${\cal AD}=\alpha=-\beta$. By $\epsilon_C=0$ this becomes the \psu\ limit of the transformation.

\section{Non-linear representation of \D}

In this section we discuss the coset space \D/\{H$\otimes$U(1)\} which non-linearly realizes the exceptional supergroup \D, whose Lie-algebra was discussed in section 2. We choose the simplest case where H=SU(2)$\otimes$SU(2).  Then 
the generators of \D\ are decomposed into the subsets denoted by 
$$
\{T_{\rm D}^\Xi\}=\{\underbrace{P,Q^a_{\ \alpha}}_{X^i},\overbrace{\underbrace{K,S^\alpha_{\ a}}_{Y^i},\underbrace{L^\alpha_{\ \beta},R^a_{\ b}}_{H^I}}^{{\hat H}^{\hat I}},C\}.  
$$
\D/\{H$\otimes$U(1)\} is parametrized by the supercoordinates $\varphi^i=(x, \theta^\alpha_{\ a})$ and the complex conjugates $\sigma^i=(y,\omega^a_{\ \alpha})$ which correspond to the generators $\{X^i\}$ and $\{Y^i\}$ respectively. 
To study this coset space, we start by considering the coset space D(2,1;$\gamma$)$^{\rm C}/\{\hat{\rm H}\otimes{\rm U(1)}^{\rm C}\}$. Here D(2,1;$\gamma$)$^{\rm C}$ is the complex extension of \D\ and $\hat{\rm H}\otimes{\rm U(1)}^{\rm C}$  is the complex subgroup generated by the generators ${\hat H}^{\hat I}$ and $C$ in the decomposition. 
Write a coset element of \D$^{\rm C}/\{\hat{\rm H}\otimes$U(1)$^C$\} as
\bb
e^{\varphi\cdot X}=e^{xP+\theta^\alpha_{\ a}Q^a_{\ \alpha}}.      \nonumber
\ee
For a left multiplication of an element $e^{i\epsilon\cdot T_{\rm D}}\in$\ \D\ the coset element changes as 
\bb
e^{i\epsilon\cdot T_{\rm D}}e^{\varphi\cdot X}e^{-i[\lambda(\varphi,\epsilon)\cdot\hat H+\lambda_C(\varphi,\epsilon)C]}=e^{\varphi'(\varphi,\epsilon)\cdot X},  \label{44}
\ee
with a compensator $e^{-i[\lambda(\varphi,\epsilon)\cdot\hat H+\lambda_C(\varphi,\epsilon)C]}$. 
 Here use was made of 
\bb
\epsilon\cdot T_{\rm D}&=& \epsilon_P P+\epsilon_Q\cdot Q+\epsilon_K K+\epsilon_S\cdot S+ \epsilon_L\cdot L+\epsilon_R\cdot R+ \epsilon_C C,   \nonumber\\
\lambda(\varphi,\epsilon)\cdot\hat H&=& \lambda_K(\varphi,\epsilon)K+\lambda_S(\varphi,\epsilon)\cdot S+\lambda_L(\varphi,\epsilon)\cdot L+\lambda_R(\varphi,\epsilon)\cdot R.  \label{44'}
\ee
(\ref{44}) defines the transformation of the coordinates $\varphi^i \rightarrow \varphi'^i(\varphi,\epsilon)$. 
When $\epsilon$ is infinitesimally small this relation defines the Killing vectors $R^{\Xi i}(\varphi)$  as
\bb
&\ & \delta\varphi^i=\epsilon^\Xi R^{\Xi i}(\varphi)=\Big(\epsilon^\Xi R^{\Xi}_P(\varphi),\epsilon^\Xi R^{\Xi\alpha}_{Q\ a}(\varphi)\Big),   \nonumber
\ee
and the parameter functions  $\lambda^\Xi(\varphi)$ and $\lambda^\Xi_C(\varphi)$ of the compensator  as 
\bb
\lambda(\varphi,\epsilon)\cdot\hat H+\lambda_C(\varphi,\epsilon)C=
\epsilon^\Xi[\lambda^\Xi(\varphi)\cdot \hat H+ \lambda^\Xi_C(\varphi)C].
 \nonumber
\ee
According to \cite{Ao5} they can be calculated in a purely algebraic way. We only outline the calculation. For the details the reader may refer to \cite{Ho}. 
For infinitesimally small parameters $\epsilon^\Xi$, we may write the transformation (\ref{44}) as 
\bb
e^{\varphi\cdot X+\epsilon^\Xi R^\Xi\cdot X+O(\epsilon^2)}=
e^{i\epsilon\cdot T_{\rm D}}e^{\varphi\cdot X}e^{-i
\epsilon^\Xi[\lambda^\Xi(\varphi)\cdot \hat H+ \lambda^\Xi_C(\varphi)C]}.  \nonumber
\ee
This becomes 
\bb
&\ & 
e^{\varphi\cdot X+\epsilon^\Xi R^\Xi\cdot X+O(\epsilon^2)}  
   \nonumber\\
&\ & \hspace{1cm} = e^{\varphi\cdot X+i\sum_{n=0}^\infty\alpha_n(ad\ \varphi\cdot X)^n(\epsilon\cdot T_{\rm D})-i\sum_{n=0}^\infty(-1)^n\alpha_n(ad\ \varphi\cdot X)^n(\epsilon^\Xi[\lambda^\Xi(\varphi)\cdot \hat H+ \lambda^\Xi_C(\varphi)C])+O(\epsilon^2)},  \label{33}
\ee
by using the following formulae: for matrices  ${\cal E}$ and $X$  
\begin{eqnarray}
\exp{\cal E}\exp X &=& \exp \biggl( X+\sum_{n=0}^\infty \alpha_n(ad\ X)^n{\cal E} +   O({\cal E}^2)   \biggr),    \label{34}  \\
\exp X\exp{\cal E} &=& \exp \biggl( X+\sum_{n=0}^\infty (-1)^n\alpha_n(ad\ X)^n{\cal E} +  
 O({\cal E}^2)   \biggr), \label{35}
\end{eqnarray}
if ${\cal E} \ll 1$. Here  $\alpha_n$ are the constants 
\begin{eqnarray}
\alpha_0 =1,\quad \alpha_1=-{1\over 2},\quad \alpha_2 ={1\over 12}, \quad
\alpha_3=0, \quad \alpha_4=-{1\over 720}, \quad\cdots\cdots.  \nonumber
\end{eqnarray}
The quantity $(ad\ X)^n$ in (\ref{34}) and (\ref{35}) is a mapping defined by the $n$-ple commutator 
\bb
(ad\ X)^n{\cal E}=[X,\cdots,[X,[X,{\cal E}]]\cdots].  \nonumber
\ee
These formulae were proved in appendix B of \cite{Ao5}.

We calculate the multiple commutators in the r.h.s. of (\ref{33}), assuming that the generators $T^\Xi$ commute with $\varphi^i$s, $\lambda^\Xi$s, $\lambda^\Xi_C$s, $\epsilon^\Xi$s irrespectively of their gradings.\footnote{For more explanation on this assumption refer to appendix B of \cite{Ao5}.}   
We then expand $R^\Xi(\varphi)$, $\lambda^\Xi(\varphi)$, $\lambda^\Xi_C(\varphi)$ in series of $\varphi^i$:
\begin{eqnarray}
R^{\Xi}(\varphi) &=& R^{\Xi}_{(0)}(\varphi) + R^{\Xi}_{(1)}(\varphi)+ \cdots + R^{\Xi}_{(n)}(\varphi)+ 
\cdots, 
\nonumber\\
\lambda^\Xi(\varphi) &=&\lambda^\Xi_{(0)}(\varphi) +\lambda^\Xi_{(1)}(\varphi)+ \cdots+ \lambda^\Xi_{(n)}(\varphi)+ \cdots,   \nonumber \\
\lambda_C^\Xi(\varphi) &=&\lambda^\Xi_{C(0)}(\varphi) +\lambda^\Xi_{C(1)}(\varphi)+ \cdots+ \lambda^\Xi_{C(n)}(\varphi)+ \cdots,  \nonumber
\end{eqnarray}
Comparing the powers of both sides of (\ref{33}) order by order yields recursive relations for $R^{\Xi}_{(n)}(\varphi),  \lambda^\Xi_{(n)}(\varphi), \lambda^\Xi_{C(n)}(\varphi)$. 
With the initial condition $\epsilon^\Xi R^{\Xi i}_{(0)}=(i\epsilon_P, i{\epsilon_Q}^\alpha_{\ a})$ we solve the relations for them. 
By lengthy calculations, but a purely algebraic use of (\ref{9})$\sim$(\ref{14}) we find  that\cite{Ho} 
\bb
-i\epsilon^\Xi R^{\Xi}_P&=& \epsilon_P+ \Big[-{\gamma\over 2}\epsilon_{\alpha\beta}\epsilon^{ab} \theta^{\alpha}_{\ a}\epsilon^{\  \beta}_{Q b} +\epsilon_C x\ \Big]+ {1\over 2}(2\epsilon_K x-\gamma\theta\epsilon_S)x    \nonumber\\
&-&{\gamma\over 12}(\alpha-\beta)\epsilon_{\alpha\beta}\epsilon^{ab}\theta^\alpha_{\ a}\theta^\gamma_{\ b}(\theta\epsilon_S)^\beta_{\ \gamma}
-{\gamma\over 24}(\alpha-\beta)\epsilon_K\epsilon_{\alpha\beta}\epsilon_{\gamma\delta}\epsilon^{ab}\epsilon^{cd}\theta^\alpha_{\ d}\theta^\beta_{\ b}\theta^\gamma_{\ c}\theta^\delta_{\ a},      \nonumber\\
-i\epsilon^\Xi R^{\Xi\alpha}_{Q\ a}&=& {\epsilon_Q}^\alpha_{\ a}+\Big[x\epsilon^{\ \ b}_{S\ \beta}
\epsilon^{\alpha\beta}\epsilon_{ab} +(\theta\epsilon_L)^\alpha_{\ a}-(\theta\epsilon_R)^{\alpha}_{\ a}+{1\over 2}\epsilon_C\theta^\alpha_{\ a} \Big]  + \epsilon_K x\theta^\alpha_{\ a}   \nonumber\\
&-&{1\over 2}\Big[(\alpha-\beta)\theta^\beta_{\ a}(\theta\epsilon_S)^\alpha_{\ \beta}-{1\over 2}[(\alpha+\beta-\gamma)\theta_{\ a}^{\alpha}(\theta\epsilon_S)\Big] \nonumber\\
&-&{1\over 6}(\alpha-\beta)\epsilon_K\epsilon_{\beta\gamma}\epsilon^{bc}\theta^\alpha_{\ b}\theta^\gamma_{\ c}\theta^\beta_{\ a},      \label{46}
\ee 
together with 
\bb
\lambda_K(\varphi,\epsilon)&=&\epsilon_K,\quad\quad 
{\lambda_{S}}(\varphi,\epsilon)^a_{\ \alpha}={\epsilon_S}^a_{\ \alpha}+\epsilon_{\alpha\beta}\epsilon^{ab}\epsilon_K \theta^\beta_{\ b},     \nonumber\\
{\lambda_{L}}^\alpha_{\ \beta}(\varphi,\epsilon)&=&{\epsilon_L}^\alpha_{\ \beta}-\alpha\Big[(\theta\epsilon_S)^\alpha_{\ \beta}-{1\over 2}\delta^\alpha_{\beta}(\theta\epsilon_S)\Big]-{\alpha\over 2}\epsilon_K\epsilon_{\beta\gamma}\epsilon^{bc}\theta^\alpha_{\ b}\theta^\gamma_{\ c},   \nonumber\\
{\lambda_{R}}^a_{\ b}(\varphi,\epsilon)&=& {\epsilon_R}^a_ {\ b}+\beta\Big[(\theta\epsilon_S)^a_{\ b}-{1\over 2}\delta^a_b(\theta\epsilon_S)\Big]+ {\beta\over 2}\epsilon_K\epsilon_{\alpha\beta}\epsilon^{ca}\theta^\beta_{\ b}\theta^\alpha_{\ c},    \nonumber\\
\lambda_{C}(\varphi,\epsilon)&=& \epsilon_C-\gamma(\theta\epsilon_S)+2x\epsilon_K.    \nonumber
\ee

So far we have discussed the coset space \D$^{\rm C}/\{\hat{\rm H}\otimes$U(1)$^{\rm C}\}$. We may enlarge the coset space to \D$^{\rm C}/\{{\rm H}\otimes$U(1)\} as
$$
{\ {\rm D}(2,1;\gamma)^{\rm C}\over {\rm H}\otimes{\rm U(1)}}={{\rm D}(2,1;\gamma)^{\rm C}\over \hat{\rm H}\otimes{\rm U(1)}^{\rm C}}
\otimes{\ \ \hat{\rm H}\otimes{\rm U(1)}^{\rm C}\over\ \ \{{\rm H}\otimes{\rm U(1)}\}^{\rm C}}\otimes{\{{\rm H}\otimes{\rm U(1)}\}^{\rm C}\over {\rm H\otimes U(1)}}
$$
Correspondingly an element $e^{i\varphi\cdot X}\in \ $\D$^{\rm C}/\{\hat{\rm H}\otimes$U(1)$^{\rm C}\}$ is generalized to
\bb
U(\varphi,\sigma)=e^{\varphi\cdot X}e^{a_Y(\varphi,\sigma)\cdot Y}e^{b_L(\varphi,\sigma)\cdot L+b_R(\varphi,\sigma)\cdot R}e^{c(\varphi,\sigma)C}\in {\ {\rm D}(2,1;\gamma)^{\rm C}\over {\rm H}\otimes{\rm U(1)} }.  
\label{U1}  
\ee 
It becomes a coset element of \D/\{H$\otimes$U(1)\} when imposed on \D\  the unitarity $U^\dagger U=UU^\dagger=1$ and the complex structure so that $X^{\dagger i}=-Y^i$ and $\varphi^{*i} = \sigma^i$. Then $a_Y(\varphi,\sigma)$ consisting of $a_K(\varphi,\sigma)$ and ${a_S}^a_{\ \alpha}(\varphi,\sigma)$ 
are complex functions, but ${b_L}^\alpha_{\ \beta}(\varphi,\sigma), {b_R}^a_{\ b}(\varphi,\sigma)$ and $c(\varphi,\sigma)$ real functions 
 by definition of the coset space $\{{\rm H}\otimes{\rm U(1)}\}^{\rm C}$/\{H$\otimes$U(1)\}\cite{Ito}. 
 They are determined by the unitarity of $U(\varphi,\sigma)$.  Practically it is done by writing the unitary condition as 
\bb
e^{-\sigma\cdot Y}e^{\varphi\cdot X}=e^{a_Y(\varphi,\sigma)^*\cdot X}e^{-2[b_L(\varphi,\sigma)\cdot L+b_R(\varphi,\sigma)\cdot R+c(\varphi,\sigma)C]}e^{-a_Y(\varphi,\sigma)\cdot Y},  \label{U2}
\ee
and evaluating both sides  in an appropriate matrix representation, say $d\times d$ supermatrix representation. 
Thus we obtain  $U(\varphi,\sigma)$ as a coset element of \D/\{H$\otimes$U(1)\}.

Now we are in a position to discuss the Killing scalar for the coset space \D/\{H$\otimes$\ U(1)\}. For a left multiplication of an element $e^{i\epsilon\cdot T_{\rm D}}\in$\ \D\ the coset element $U(\varphi,\sigma)$ transforms as
\bb
U(\varphi,\sigma)\longrightarrow e^{i\epsilon\cdot T_{\rm D}} U(\varphi,\sigma)
 e^{-i\rho(\varphi,\sigma,\epsilon)\cdot H-i\rho_C(\varphi,\sigma,\epsilon)C}=U(\varphi'(\varphi,\epsilon),\sigma'(\sigma,\epsilon)),   \label{U3}
\ee
with an appropriately chosen compensator $e^{i\rho(\varphi,\sigma,\epsilon)\cdot H+i\rho_C(\varphi,\sigma,\epsilon)C}$. We have assumed a $d\times d$ supermatrix representation for $U(\varphi,\sigma)$. 
Then the Killing scalar $\Upsilon(\varphi,\sigma)$ of the coset space \D/\{H$\otimes$U(1)\}, which transforms as 
\bb
\Upsilon(\varphi,\sigma)\longrightarrow e^{i\epsilon\cdot T_{\rm D}}\Upsilon(\varphi,\sigma),    \label{KStransf}
\ee
is given by taking any  column vector from $U(\varphi,\sigma)$, for example  
\bb
\Upsilon(\varphi,\sigma)=\sum_{d=1}^n\left(
\begin{array}{c}
U^1_{\ d}(\varphi,\sigma) \\
U^2_{\ d}(\varphi,\sigma)  \\
 \vdots   \\             
U^n_{\ d}(\varphi,\sigma) 
\end{array}\right)\eta^d.      \label{KS}
\ee
Here $\eta$ is a quantity transforming  as $\eta\rightarrow e^{i\rho(\varphi,\sigma,\epsilon)\cdot H+i\rho_C(\varphi,\sigma,\epsilon)C}\eta$, which is induced  by the transformation (\ref{U3}). Its existence was shown in \cite{Ao2}.

\section{Symmetry reduction to \psu\U}

Now we come to the main point of this letter. Let us reduce the \D\ symmetry to \psu\U\  in the non-linear representation. 
The Lie-algebra in this limit was given (\ref{11})$-$(\ref{16}), while (\ref{11})$-$(\ref{13}) vanishing after the rescaling $(C,P,K)\rightarrow {1\over \gamma}(C,P,K)$.  To find the Killing vectors (\ref{46}) in the scaling limit, the calculation in the previous section should be redone with these algebra. We then find  them to tend to 
\bb
-i\epsilon^\Xi R^{\Xi}_P&=& \epsilon_P-{1\over 2}\epsilon_{\alpha\beta}\epsilon^{ab} \theta^{\alpha}_{\ a}\epsilon^{\  \beta}_{Q b}   
-{1\over 12}(\alpha-\beta)\epsilon_{\alpha\beta}\epsilon^{ab}\theta^\alpha_{\ a}\theta^\gamma_{\ b}(\theta\epsilon_S)^\beta_{\ \gamma},       \nonumber\\
-i\epsilon^\Xi R^{\Xi\alpha}_{Q\ a}&=& {\epsilon_Q}^\alpha_{\ a} +(\theta\epsilon_L)^\alpha_{\ a}-(\theta\epsilon_R)^{\alpha}_{\ a}  
-{1\over 2}(\alpha-\beta)\theta^\beta_{\ a}(\theta\epsilon_S)^\alpha_{\ \beta}.   \label{46'}
\ee 
It amounts to dropping  the coupling terms of the parameters $\epsilon_C,\epsilon_P,\epsilon_K$ and the coordinate $x$ in (\ref{46}). It is because the multiple commutators with $C,P,K$ in the r.h.s. of (\ref{33})
 are vanishing when calculated  in the limit. 
Correspondingly the coset space \D/\{H$\otimes$U(1)\} is reduced to \psu\U/\ \{H$\otimes$U(1)\}. The transformation (\ref{46'})  represents the \psu\U\ symmetry on this reduced coset space.  
Thus a non-simple group symmetry such as \psu\U\ has been non-linearly realized  in a definite way. Without the above symmetry reduction  it would  have been done hardly. This is an important point in this letter. Thus 
 the coset space \psu\U/\{H$\otimes$U(1)\} gets well-defined by the Killing vectors (\ref{46'}).   For this coset space we may find   the Killing scalar as well.  Here also we had better recalculate it following the procedure (\ref{U1})$-$(\ref{KS}), rather than resort to a scaling argument of (\ref{KS}). It takes the form 
\bb
\Upsilon_{{\rm PSU}\otimes{\rm U(1)}^3}(\varphi,\sigma) =e^{x P+y K}\Upsilon_{\rm PSU}(\theta,\omega),
\label{49}
\ee
with $\Upsilon_{\rm PSU}(\theta,\omega)$ the Killing scalar for the coset space \psu/H. The phase factor is due to the decoupling of $C,P,K$ in the limit. 
 It is not given in the real basis, but a U(1) factor due to the unitary condition $P^\dagger=-K$ and $x^*=y$, explained below (\ref{U1}).
 This form of the Killing scalar 
  may be alternatively understood  by writing  the coset space \psu\U/\{H$\otimes$U(1)\}  as \psu$\otimes$U(1)$^2$/H. Namely  the phase factor in (\ref{49}) is due to the U(1)$^2$ charge.  
The Killing scalar (\ref{49}) transforms as 
\bb
\Upsilon_{{\rm PSU}\otimes{\rm U(1)}^3}(\varphi,\sigma)  \longrightarrow e^{i\epsilon\cdot T_
{{\rm PSU}\otimes{\rm U(1)}^3}}\Upsilon_{{\rm PSU}\otimes{\rm U(1)}^3}(\varphi,\sigma),  \label{KStransf'}
\ee
by the non-linear transformation (\ref{46'}) by  construction. It is interesting to observe that  the linear transformation (\ref{KStransf'}) is  obtained  although the phase factor $e^{x P+y K}$ is subjected to the non-linear transformation by the Killing vector $R^\Xi_P$  and its complex conjugate $R^\Xi_K$  in (\ref{46'}). 
(\ref{KStransf'}) is isomorphic to  the transformation (\ref{22}). Hence $\Upsilon_{{\rm PSU}\otimes{\rm U(1)}^3}$ may be identified with the spin-variable $\psi_m$ of the spin-system with \psu\U. Consequently we are led to claim that the non-linear $\sigma$ model on \psu\U/\{H$\otimes$U(1)\} is equivalent to the spin-system with the centrally extended symmetry \psu\U.

In the \psu\U\ limit the assumed $d\times d$ matrix representation  in (\ref{KS}) becomes  reducible to that of 4$\times$4 matrix. When we use the latter representation the equivalence between the non-linear $\sigma$-model and the spin-system becomes more clear, by writing the Killing scalar $\Upsilon_{\rm PSU}(\theta,\omega)$  in (\ref{49}) in an explicit form. 
To this end we need an element of \psu/H similar to (\ref{U1}), 

 \bb
U(\theta,\omega)=e^{\theta\cdot Q}e^{a(\theta,\omega)\cdot S}e^{b_L(\theta,\omega)\cdot L+b_R(\theta,\omega)\cdot R+c(\theta,\omega)C} \in
 {\rm PSU}(2|2)/{\rm H}.    \label{39}
\ee
The parameter functions  are obtained as the \psu\U\ limit of $U(\varphi,\sigma)\in$ \D/\ \ \{H$\otimes$U(1)\}, but may be directly found by repeating the procedure again. We work out the unitary condition (\ref{U2}) for this case. 
 Use the matrix representation (\ref{20}) of \psu\U\ in the \psu\ limit, defined by (\ref{19}). 
 We then find (\ref{39})  to be  
\bb
U(\theta,\omega)= \left(
\begin{array}{c|c}
U^\gamma_{\ \delta}(\theta,\omega)& U^\gamma_{\ d}(\theta,\omega) \\
\vspace{-0.3cm}   \\
\hline
\vspace{-0.3cm}   \\
U^c_{\ \delta}(\theta,\omega) & U^c_{\ d}(\theta,\omega) 
\end{array}\right)=
\left(
\begin{array}{c|c}
\hspace{0.1cm}{1\over \sqrt{1+\theta\omega}}\hspace{0.1cm} &\hspace{0.1cm} {{\cal A}\over \sqrt{1+\theta\omega}}\theta\hspace{0.3cm}  \\
\vspace{-0.3cm}  &    \\
\hline 
\vspace{-0.3cm}  &    \\
\hspace{0.1cm}{-{\cal D}\over \sqrt{1+\omega\theta}}\omega\hspace{0.1cm} &  {1\over \sqrt{1+\omega\theta}}
\end{array}\right).    \nonumber
\ee
Here note that ${\cal A}^*={\cal D}$ since we have the constraint ${\cal AD}=\alpha=-\beta$ from (\ref{21}).  Hence the complex structure was assumed also for \psu. With this $U(\theta,\omega)$ 
  the Killing scalar $\Upsilon_{\rm PSU}(\theta,\omega)$ of the form (\ref{KS}) becomes explicit as
\bb
\Upsilon_{\rm PSU}(\theta,\omega)=\left(
\begin{array}{c}
U^\gamma_{\ d}(\theta,\omega)\\
\vspace{-0.3cm}   \\
\hline 
\vspace{-0.3cm}   \\
U^c_{\ d}(\theta,\omega)
\end{array}\right)\eta^d
=
\left(
\begin{array}{c}
 {{\cal A}\over \sqrt{1+\theta\omega}} \theta \\
\vspace{-0.3cm}     \\
\hline 
\vspace{-0.3cm}     \\
{1\over \sqrt{1+\omega\theta}}
\end{array}\right)\eta,     \label{41}
\ee
for instance. Thus the Killing scalar $\Upsilon_{{\rm PSU}\otimes{\rm U(1)}^3}(\varphi,\sigma)$ in (\ref{49}) represents two-component spinors of the spin-variable $\psi_m$.

\section{Conclusion}

In this letter we have discussed the central extension  of \psu\ to \psu\U\ in the non-linear representation. To this end  we  studied the coset space \D/\{H$\otimes$U(1)\} choosing H to be SU(2$|$2)$\otimes$SU(2$|$2). Then the coset space \psu\U/\{H$\otimes$U(1)\} was well-defined by the symmetry reduction of \D\ to \psu\U\ in the former coset space. The Killing scalar $\Upsilon_{{\rm PSU}\otimes{\rm U(1)}^3}(\varphi,\sigma)$ for this reduced coset space was obtained  in the form (\ref{49}).  It was shown to transform as (\ref{KStransf'}) by the non-linear transformation  of the reduced coset space, given by (\ref{46'}). 
 Identifying $\Upsilon_{{\rm PSU}\otimes{\rm U(1)}^3}(\varphi,\sigma)$ with the spin-variable $\psi_m$ we have claimed that the non-linear $\sigma$-model on  \psu\U/\{H$\otimes$U(1)\} is equivalent to the spin-system with the \psu\U\ symmetry. 

The central extension  of \psu\ to \psu\U\ in the non-linear representation can be similarly done by choosing H to be U(1)$\otimes$U(1) instead of SU(2$|$2)$\otimes$SU(2$|$2).  For this case the \psu\U\ symmetry is  realized on the coset space \psu\U/U(1)$^3$, which is a central extension of the coset space \psu/U(1)$^2$. It is also obtained from \D/U(1)$^3$ by the symmetry reduction discussed in this letter. To study these coset spaces we have to furthermore decompose the generators  of \psu\ as
$$
\{T^\Xi_{\rm PSU}\}=\{\underbrace{L^1_{\ 2},L^2_{\ 1},R^1_{\ 2},R^2_{\ 1}}_{\cal P},\underbrace{Q^a_{\ \alpha},S^\alpha_{\ a}}_{\cal Q}, \underbrace{L^1_{\ 1}(=-L^2_{\ 2}),R^1_{\ 1}(=-R^2_{\ 2})}_{\cal H}\}. 
$$
The coset space \psu/U(1)$^2$ is mimic to PSU(2,2$|$4)/\{SO(1,4)$\otimes$SO(5)\} discussed in \cite{Tsey}. It is parametrized by the coordinates $X_{{\rm S}^2\otimes{\rm S}^2}$ of S$^2\otimes$S$^2$ and the fermionic ones $\Theta$, which correspond to the coset generators $\cal P$ and $\cal Q$ respectively. The Killing scalar for the coset space  takes the same form as for PSU(2,2$|$4)/\{SO(1,4)$\otimes$SO(5)\}\cite{Ao5}
\bb
\Upsilon_{\rm PSU}(X,\Theta)=e^{X_{{\rm S}^2\otimes{\rm S}^2}\cdot {\cal P}+ \Theta  \cdot \cal Q}\eta.    \label{50}
\ee
It resembles to the vertex operator of the Green-Schwarz superstring. We consider the enlarged coset space  \D/U(1)$^3$ by introducing the coordinates $x$ and $y(=x^*)$ as previously. For this  enlarged coset space there exists the Killing scalar, say $\Upsilon(x,y,X_{{\rm S}^2\otimes{\rm S}^2},\Theta)$. The central extension of the Killing scalar (\ref{50}) is the limit of it in which the enlarged coset space \D/U(1)$^3$ gets reduced to \psu\U/U(1)$^3$, i.e.,
\bb
\Upsilon_{{\rm PSU}\otimes{\rm U(1)}^3}(x,y,X_{{\rm S}^2\otimes{\rm S}^2},\Theta) =e^{x P+y K}\Upsilon_{\rm PSU}(X_{{\rm S}^2\otimes{\rm S}^2},\Theta).
  \label{51}
\ee
It corresponds to (\ref{49}) in the previous argument, and has the same transformation property as (\ref{KStransf'}) by the non-linear transformation realized on the \psu\U/U(1)$^3$.

Either of the Killing scalars $\Upsilon(\varphi,\sigma)$ and $\Upsilon(x,y,X_{{\rm S}^2\otimes{\rm S}^2},\Theta)$, given in a form such as (\ref{KS}), 
satisfies the exchange algebra with the universal R-matrix of \D, when non-linear $\sigma$-models on those coset spaces are quantized following \cite{Ao2,Ao5}. The universal R-matrix becomes position-dependent as the consequence of the symmetry reducing of these Killing scalars as (\ref{49}) and (\ref{51}) respectively. 
The real problem is to understand how the U(1) phase factors in (\ref{49}) and (\ref{51}) are braided in the correlation functions (\ref{cor}) so that the R-matrix satisfies the Yang-Baxter equation\cite{Bei}. We hope that the non-linear representation presented in this letter would shed new light on such a study.

\end{document}